\documentclass[11pt]{article}

\usepackage{a4wide}
\usepackage{latexsym}
\usepackage{graphicx}
\usepackage{color}
\usepackage{hyperref}

\title{A Two-Player Game of Life}

\author{Mark Levene and George Roussos \\
School of Computer Science and Information Systems \\
Birkbeck College, University of London \\
London WC1E 7HX, U.K. \\ \{mark,gr\}@dcs.bbk.ac.uk}

\date{}

\begin{document}

\maketitle

\begin{abstract}

We present a new extension of Conway's game of life for two players,
which we call {\em p2life}. P2life allows one of two types of token,
black or white, to inhabit a cell, and adds competitive elements into the
birth and survival rules of the original game. We solve the mean-field
equation for p2life and determine by simulation that the asymptotic
density of p2life approaches $0.0362$.

\end{abstract}
\medskip


\section{Introduction}

Conway's game of life \cite{BERL82} is the best known two-dimensional
cellular automaton \cite{TOFF87}, due to the complex behaviour it
generates from a simple set of rules. The game of life, introduced to the
world at large by Martin Gardner in 1970 \cite{GARD70}, has provided
throughout the years challenging problems to many enthusiasts. A large
number of these are documented in Stephen Silver's comprehensive life
lexicon \cite{SILV02}. Detailed accounts of the game of life have been
given by Poundstone in \cite{POUN85}, which uses the game of life as an
illustration of complexity, and by Sigmund in \cite{SIGM93}, which uses
the game of life in the context of artificial life and the ability of a
cellular automaton to self-replicate. An interesting extension of the
game of life to three dimensions was suggested by Bays \cite{BAYS87}.

\medskip

The game of life is played on a square lattice with interactions to
nearest and to next-nearest neighbours, where each cell can be either
empty or occupied by a token and is surrounded by eight neighbouring
cells. The evolution of the game is governed by the following simple
rules. If a cell is empty it gives {\em birth} to a token if exactly
three of its neighbours are occupied, and if it is occupied it {\em
survives} if either two or three of its neighbours are occupied. In all
other cases either the cell remains empty or it {\em dies}, i.e. it
becomes empty. The game evolves by repeated applications of these rules
to produce further configurations. The single player of the game decides
what the initial configuration of the lattice will be and then watches
the game evolve. One of the questions that researchers have investigated
is the asymptotic behaviour of this evolutionary process. It transpires
that when the initial configuration is random and its density is high
enough, then the game eventually stabilises to a density of about 0.0287
\cite{BAGN91,GIBB97}; see also \cite{MALA98}.

\medskip

One aspect that is missing from Conway's game of life is the
competitiveness element of two-player games, as Gardner noted ``Attempts
have also been made to invent competitive games based on ``Life'', for
two or more players, but so far without memorable results.'' \cite[p.
255]{GARD83}. This is the challenge that we take up in this paper.

\smallskip

An interesting version of the game of life for two players is known as
{\em black and white} or {\em immigration} \cite{SILV02}, where cells are
either white or black and when a birth occurs the colour of the token is
decided according to the majority of neighbouring cells. Although this
variation is interesting in its own right, survival does not involve the
two colours and remains non-competitive as in the single player version.
We propose a new two-player version of the game of life where both birth
and survival are competitive, and provide a preliminary analysis of its
behaviour.

\medskip

The rest of the paper is organised as follows. In Section~\ref{sec:rules}
we present the rule set for our two-player version of the game of life
and in Section~\ref{sec:mean-field} we provide a mean-field analysis of
the game. In Section~\ref{sec:density} we present results of simulations
to ascertain the asymptotic density of the game and, finally, in
Section~\ref{sec:conc} we give our concluding remarks and provide a web
link to an implementation of the game.

\section{Rules of the Game}
\label{sec:rules}


In the two-player variation of the game of life, which we call {\em
p2life}, the players, white and black, are competing for space. Conway's
game of life is considered to be ``interesting'' since its simple set of
rules lead to complex and unpredictable behaviour. (The notion that from
simple rules complex behaviour can emerge, which may help us understand
the diversity of natural phenomena, is discussed in great detail by
Wolfram in \cite{WOLF02} but can already be learned from the 130 year old
thesis of van der Waals  on liquid-vapour equilibria.) P2life maintains
the ``interesting'' behaviour of the game of life by preserving the
essence of Conway's game and adding to its rules a competitive element to
decide who will give birth and who will survive.

\medskip

The rules of p2life, from white's point of view (the rules from black's
point of view are symmetric), are as follows:
\begin{description}
\item[Birth.] If a cell is empty, then we consider two cases:
\renewcommand{\labelenumi}{\arabic{enumi})}
\begin{enumerate}
\item The cell has exactly three white neighbours and the number of black
neighbours is different from three. In this case a white token is born in
the cell.

\item The cell has exactly three white and three black neighbours. In
this case an unbiased coin determines whether a white or black token is
born in the cell.
\end{enumerate}

\item[Survival.] If a cell is occupied by a white token, then we consider
two cases:
\renewcommand{\labelenumi}{\arabic{enumi})}
\begin{enumerate}
\item If the difference between the number of white and black neighbours
is two or three, then the white token survives.

\item If the difference between the number of white and black neighbours
is one and the number of white neighbours is at least two, then the white
token survives.
\end{enumerate}
\end{description}
\medskip

It is clear that if there is only one colour on the lattice then p2life
reduces to the standard one-player version of the game. We note that
having non-symmetric rules for white and black would allow us to
investigate how different behaviour sets interact, but finding such a set
of rules which would be ``interesting'' is an open problem. During the
process of deciding the rule set for p2life we considered several
variations, which we now briefly discuss:

\renewcommand{\labelenumi}{\arabic{enumi})}
\begin{enumerate}
\item In the first part of the birth rule, insisting that the number of
black tokens should be less than three, which is similar to birth in  the
black and white variant. This decreases the birth rate but still seems to
be ``interesting''.

\item Omitting the second part of the birth rule, i.e. the random choice
when there are exactly three white and three black neighbours. Again,
this decreases the birth rate but still seems to be ``interesting''.

\item Omitting the second part of the survival rule, i.e. survival when
the difference between the number of white and black tokens is one and
the number of white tokens is at least two.  This decreases the survival
rate and the conflict between the two players but, once again seems
``interesting''

\item In the second part of the survival rule omitting the side condition
that the number of white neighbours must be at least two, or in the first
part of the survival rule omitting the condition that the difference
between white and black is at most three. Due to the increased survival
rates, these modifications lead to steady growth with spatial boundaries
between the two players. It would be interesting to compare these rules
to Schelling's models of segregation \cite{SCHE71} or the Ising model
\cite{SPIR01}.

\end{enumerate}
\medskip

We illustrate two configurations which lead to interesting confrontations
between white and black. The configuration shown on the left-hand side of
Table~\ref{table:gol1} leads to a black block and two white gliders as
shown on the right-hand side of the table. In the one-player game of life
this initial configuration annihilates itself into empty space. On the
other hand, the configuration shown on the left-hand side of
Table~\ref{table:gol2} leads to two black blocks as shown on the
right-hand side of the table. Thus black wins from this position. In the
one-player game of life this initial configuration leads to six blinkers.

\begin{table}[ht]
\begin{center}
\begin{tabular}{|c|c|ccc|c|c|c|c|c|c|c|c|c|c|c|c|} \cline{1-2}
 $\bullet$ & $\bullet$ & & & \multicolumn{12}{c}{} \\ \cline{1-2}
 $\bullet$ & $\bullet$ & & & \multicolumn{12}{c}{} \\ \cline{1-2} \cline{6-17}
 $\circ$ & $\circ$     & & & & & $\circ$ & & & & $\bullet$ & $\bullet$ & & & & $\circ$ & \\ \cline{1-2} \cline{6-17}
 $\circ$ & $\circ$     & & $\Longrightarrow$ & & $\circ$ & $\circ$ & & & & $\bullet$ & $\bullet$ & & & & $\circ$ & $\circ$  \\ \cline{1-2} \cline{6-17}
 $\circ$ & $\circ$     & & & & $\circ$ & & $\circ$ & & & & & & & $\circ$ & & $\circ$ \\ \cline{1-2}\cline{6-17}
 $\circ$ & $\circ$     & & \multicolumn{12}{c}{} \\ \cline{1-2}
 $\circ$ & $\circ$     & &  \multicolumn{12}{c}{} \\ \cline{1-2}
\end{tabular}
\end{center}
\caption{\label{table:gol1} An interesting configuration}
\end{table}
\medskip

\begin{table}[ht]
\begin{center}
\begin{tabular}{|c|c|c|c|c|c|ccc|c|c|c|c|c|c|} \cline{1-6} \cline{10-15}
 $\bullet$ & $\bullet$ & $\bullet$ & $\bullet$ & $\bullet$ & $\bullet$ & & & & $\bullet$ & $\bullet$ & & & $\bullet$ & $\bullet$ \\ \cline{1-6} \cline{10-15}
 & & $\circ$ & $\circ$ & & & & $\Longrightarrow$ & & $\bullet$ & $\bullet$ & & & $\bullet$ & $\bullet$  \\ \cline{1-6} \cline{10-15}
 & & $\circ$ & $\circ$ & & & &  \\ \cline{1-6}
\end{tabular}
\end{center}
\caption{\label{table:gol2} Another interesting configuration}
\end{table}
\medskip

\section{Mean-Field Theory for P2life}
\label{sec:mean-field}

To compute the initial configuration given a specified initial density of
$p$, we use the following procedure for each cell in the lattice. Firstly
we determine whether it is occupied or empty according to the given
probability $p$ and then, if it should be occupied we toss an unbiased
coin to decide whether it will be occupied by a white or black token. In
a similar fashion to the game of life \cite{BAGN91}, we now use
mean-field theory to determine the density of the tokens after applying
the rules to the initial configuration.

\medskip

It can be verified by the rules of p2life that the density $p^\prime$,
after a single application of the rules to a square lattice having
initial density $p$, is given the mean-field equation,
\begin{equation}\label{eq:mean-field}
p^\prime = 21\,{p}^{3}-63\,{p}^{4}+{\frac
{105}{2}}\,{p}^{5}+35\,{p}^{6}-{\frac {1505}{16}}\,{p}^{7}+{\frac
{1057}{16}}\,{p}^{8}-{\frac {553}{32}}\,{p}^{9}.
\end{equation}
\smallskip

\begin{figure}[ht]
\centerline{\includegraphics[width=12cm,height=9.33cm]{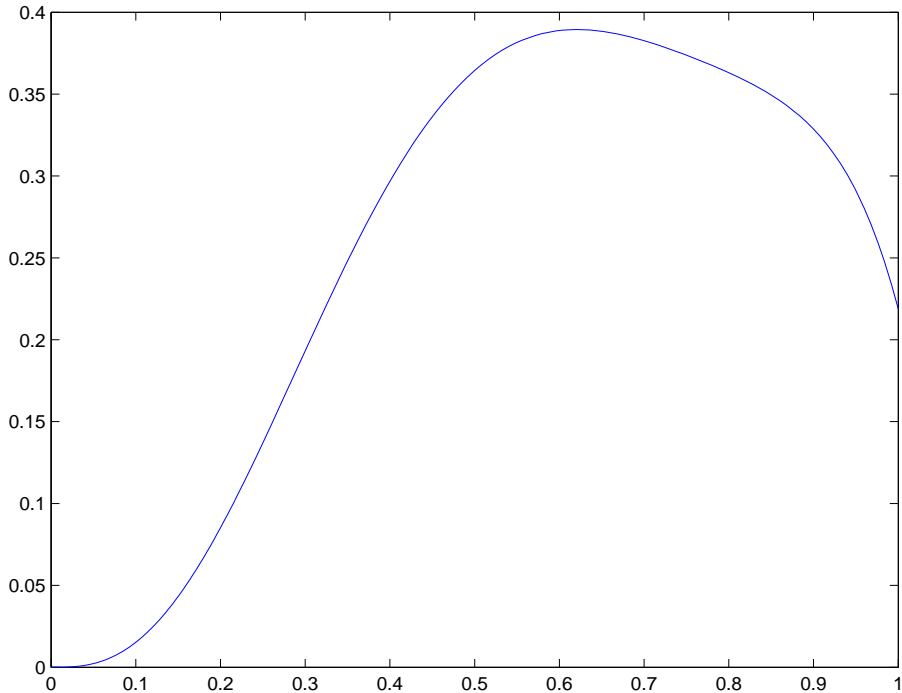}}
\caption{\label{fig:mean-field} Plot of the mean-field density $p^\prime$
against the initial density $p$}
\end{figure}

From (\ref{eq:mean-field}) we can compute the maximum density of
$p^\prime$, which is $0.3895$ at an initial density of $p = 0.6206$. This
can also be seen from the plot of the mean-field equation shown in
Figure~\ref{fig:mean-field}. As we would expect, simulations show a very
close match with this plot, although obviously due to correlations we
cannot use the mean-field equation to predict the long-term behaviour of
the game. An interesting point to note about the plot is that when $p =
1$, $p^\prime = 0.2188$, which is different from the original game of
life (and the black and white variant) where $p^\prime = 0$. The reason
is that although dense regions of a single colour die off immediately,
mixed spaces between the two colours allow for survival of tokens. We
also note that from Figure~\ref{fig:mean-field} we can verify that the
only fixed-point of the mean-field equation for p2life is zero. To
improve the prediction power for repeated applications of the rules we
could extend the mean-field analysis using the local structure theory of
Gutowitz and Victor \cite{GUTO87}.

\medskip

\section{Asymptotic Density for P2life}
\label{sec:density}


We have investigated the properties of p2life with particular emphasis on
the estimation of its asymptotic density via simulations using
\textsc{matlab}. This approach offers flexibility and reasonable
performance (0.7 million updates per second) for a particularly
computationally intensive task. We have performed simulations on both
periodic (toroidal) and cutoff boundary conditions on square lattices of
sizes from $10\times 10$ to $1000 \times 1000$. In particular, we have
investigated the dependence of the asymptotic density $p_\infty$ on the
initial density $p$ of a random, uniformly and independently distributed,
initial configuration. Each configuration is iterated until the game
configuration reaches a stable or oscillatory state. In general, it is
straightforward to detect when such conditions occur, with the notable
exception of the case of periodic boundary conditions when in the final
configuration there exists a glider which travels around the torus
without colliding with other tokens in occupied cells.

\smallskip

\begin{figure}[ht]
\centerline{\includegraphics[width=12cm,height=9.33cm]{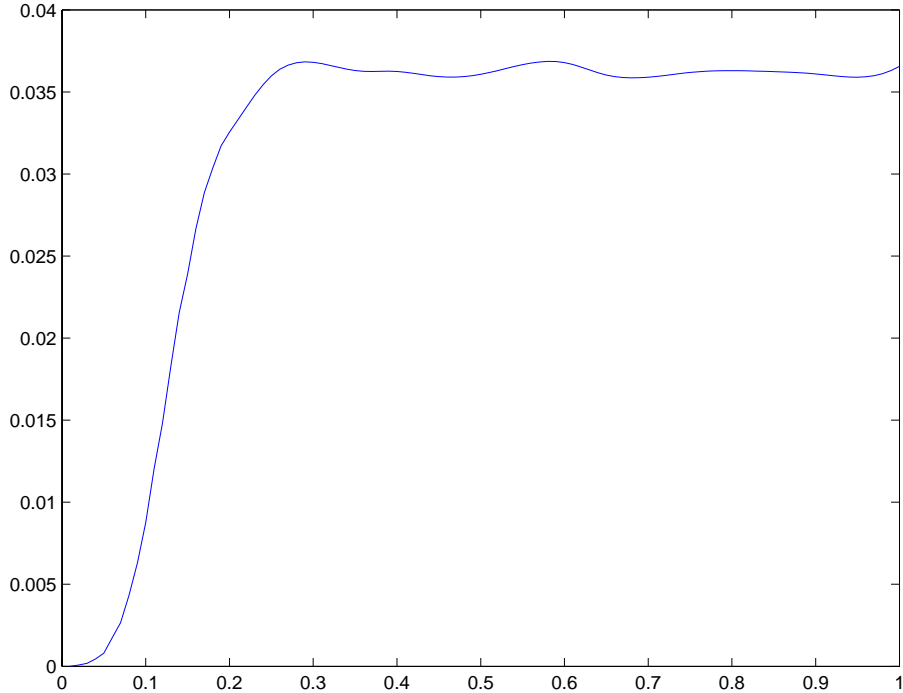}}
\caption{\label{fig:asymptotic} Plot of asymptotic density $p_\infty$
against the initial density $p$}
\end{figure}

The results for a square lattice of size $100 \times 100$ are shown in
Figure~\ref{fig:asymptotic}. Each experimentally computed point plotted
in Figure~\ref{fig:asymptotic} is the average over one hundred
independently selected initial configurations. It is immediately evident
that in direct contrast to the one-player version of the game of life,
the asymptotic density of p2life is non-zero for non-zero initial
density. In the one-player version of the game, if the initial density
increases above approximately $0.85$ the asymptotic density is zero due
to the annihilation of all tokens during the first iteration due to
overcrowding. In fact, for $p=1$ we estimate that p2life has asymptotic
density $p_\infty=0.0362$. This is due to the fact that in p2life the
survival rules allow members of both token populations to carry over to
the next iteration with $p^\prime = 0.2188$, which is consistent with the
value obtained via the mean-field approach.

\smallskip

\begin{figure}[ht]
\centerline{\includegraphics[width=12cm,height=9.33cm]{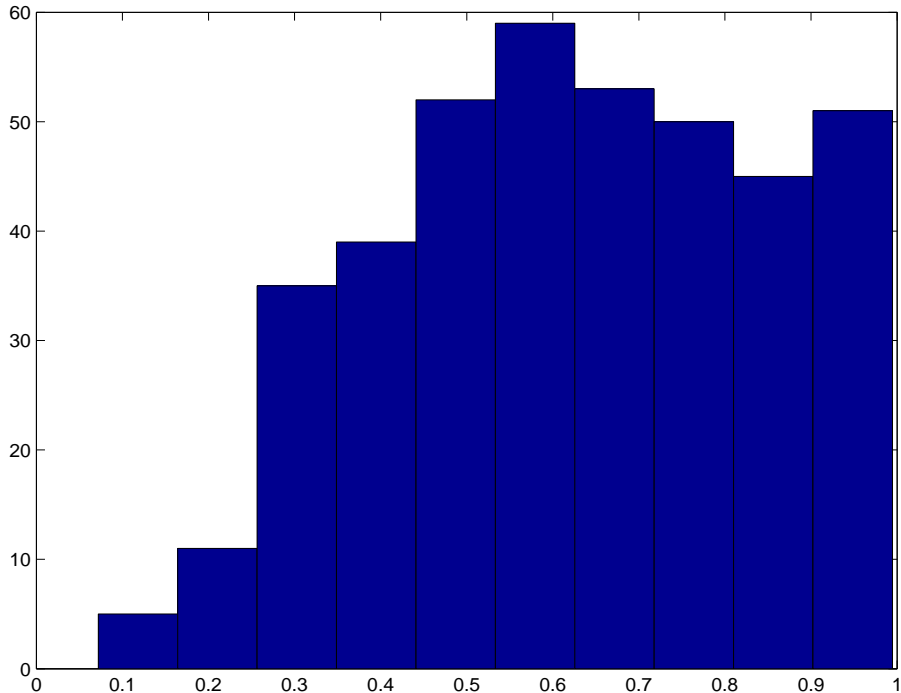}}
\caption{\label{fig:ratio}Histogram for the ratio of the loser population
over the winner population}
\end{figure}

We make several other interesting observations on the asymptotic
behaviour of p2life. Firstly, we have estimated that the maximum
asymptotic density of $p_\infty=0.0362$ is reached at about $p=0.3$ and
remains fairly constant until $p=1$. This is different to the behaviour
of the one-player version of the game, where a fairly constant asymptotic
density $p_\infty=0.0287$ is reached at about $p=0.15$ and remains at
that level until approximately $p=0.70$ \cite{BAGN91}. Secondly, when
periodic boundary conditions are used, the asymptotic density $p_\infty$
increases by approximately 5\% to $0.0381$ compared to the situation of
cutoff boundary conditions. Thirdly, it appears that the size of the
lattice does not seem to affect the asymptotic density, although, as in
the one-player version there may be small finite-size effects
\cite{GIBB97}. Finally, we estimated the ratio of the loser population
over the winner population at the final state. A histogram of the results
aggregated over 400 runs for a square lattice of size $100 \times 100$,
where  100 runs were carried out for initial densities of 0.25, 0.5, 0.75
and 1.00, is shown in Figure~\ref{fig:ratio}. We observe that over 69\%
of the runs resulted in the ratio being over 0.5, i.e. the loser having
more that one third of the final population.

\medskip

\medskip

\section{Concluding Remarks}
\label{sec:conc}

Our main contribution is to have shown that, by injecting competitive
elements into Conway's game of life, ``Life'' can be ``interesting'' when
more than one player participates in the game. An applet demonstrating
p2life can be accessed at
\begin{center}
\href{http://hades.dcs.bbk.ac.uk/users/gr/p2life/p2life.asp}{http://hades.dcs.bbk.ac.uk/users/gr/p2life/p2life.asp}.
\end{center}
\smallskip

A problem that we are now investigating is how to convert p2life into a
``real'' game, i.e. where players are allowed to make moves between
generations, which change the configuration of the tokens, and devise
strategies to overpower or live side-by-side with their opponent.


\end{document}